\documentstyle[amssymb,prb,aps]{revtex}

\begin{document}
\title{Memory of multiple aging stages above the freezing temperature in the
relaxor ferroelectric PLZT}
\author{F. Cordero$^{1}$, F. Craciun$^{1}$, A. Franco$^{1}$, D. Piazza$^{2}$ and C.
Galassi$^2$}
\address{$^{1}$ CNR, Istituto di Acustica and Istituto dei Sistemi Complessi,\\
Area della Ricerca di Roma-Tor Vergata,Via del Fosso del Cavaliere\\
100,I-00133 Roma, Italy}
\address{$^{2}$ CNR-ISTEC, Via Granarolo 64, I-48018 Faenza, Italy}
\maketitle

\begin{abstract}
The dynamic dielectric susceptibility and elastic compliance of the relaxor
ferroelectric PLZT 9/65/35 have been measured under different cooling and
heating protocols, in order to study aging and memory. The memory of
multiple aging stages at different temperatures has been found (several dips
in the susceptibility curves on heating), as in spin glass systems below the
glass transition. Remarkably, in PLZT the memory of several aging stages is
retained also above the freezing temperature deduced from the dynamic
susceptibilities. The results are discussed in the light of the existing
models of aging and memory in spin and dipolar glasses.
\end{abstract}

\twocolumn


Aging, rejuvenation and memory are considered as manifestations of the
frozen spin glass state. In the case of the dynamic susceptibility $\chi
\left( \omega ,T\right) $, aging consists in a slow reduction of $\chi $
during a stay at a constant temperature $T_{1}$ {\it below} the glass
transition temperature $T_{g}$.\cite{NMN00,DVB01,MPP01} On further cooling, $%
\chi $ may recover the values found before aging, rejoining the reference
curve that is measured during continuous cooling; this is called
rejuvenation, since the susceptibility behaves as if aging at $T_{1}$ did
not occur. On subsequent heating through $T_{1}$, $\chi $ may completely or
partially retrace the dip formed during aging, and this is called memory. A
peculiarity of spin glasses is that aging stages at several $T_{n}<T_{g}$
may be recalled, giving rise to susceptibility curves measured on continuous
heating with dips at the various $T_{n}$.\cite{BDH01} Tentative explanations
of this phenomenon involve the concept of a hierarchy in the potential
landscape of the metastable states as a function of temperature, below the
glass transition temperature.\cite{VHO96}

Relaxor ferroelectrics are often considered among the possible realizations
of spin glasses, with electric instead of magnetic dipoles, since they
exhibit most of the phenomena characterizing spin glasses, including aging,
rejuvenation and memory.\cite{CCW01,KB02b} These materials, mostly of the
perovskite type, have substantial disorder in the valence and/or size of the
cation sublattices, so that long range order in the electric polarization is
never established (unless a bias field above a certain threshold is
applied). Below the so-called Burns temperature $T_{B}$, which corresponds
to the Curie temperature $T_{{\rm C}}$ for a solid solution with a pure
ferroelectric, fluctuating polar clusters of nanometer size start forming.
On cooling at much lower temperature,\cite{KBP00} the freezing of these
polar clusters gives rise to the typical frequency dispersion in the ac
susceptibility, as found in spin glasses.

Many properties of the relaxor ferroelectrics may be explained in terms of
the recently proposed spherical random-bond random-field model,\cite{PB99}
reminiscent of the Sherrington and Kirkpatrick model of spin glasses;\cite
{SK75} the latter is the basis for various theoretical treatments of aging
and memory.\cite{MPP01,LOH91}

The material under study here is (Pb/La)(Zr/Ti)O$_{3}$ with La/Pb ratio $%
x=0.09$ and Ti/Zr ratio $y=0.35$ (PLZT $x/65/35$). The parent PZT 65/35 is a
normal ferroelectric below $T_{{\rm C}}=627$~K.\cite{VXP93} The partial
substitution of Pb with La induces the formation of the polar nanoclusters
below $T_{B}\simeq $ $T_{{\rm C}}($PZT 65/35$)\simeq 627$~K, and the relaxor
behavior with the frequency dispersive maximum of the dielectric
susceptibility around 340~K for $x=0.09$.

Here we show that in PLZT $9/65/35$ the memory of multiple aging stages,
which is generally considered peculiar of the frozen spin glass state, is
found even above the freezing temperature indicated by both the dielectric
and elastic susceptibilities.


The ceramic material was prepared by solid state reaction of the starting
oxides according to the formula Pb$_{1-x}$La$_{x}$(Zr$_{0.65}$Ti$_{0.35}$)$%
_{1-x/4}$O$_{3}$ ($x=0.09$)\ with the vacancies compensating La$^{3+}$\ for
Pb$^{2+}$\ in the Zr/Ti sublattice. The oxide powders were calcined for 4 h
at 850~$^{{\rm o}}$C, and sintered at 1200~$^{{\rm o}}$C for 2 h and at 1300~%
$^{{\rm o}}$C for 2 h. The density was 97\% of the nominal value and the
mean grain size was about 3 $\mu $m. The structure determined by X-ray
diffraction was pure perovskite without any pyrochlore phase. The samples
were cut into bars $45\times 4\times 0.5$~mm, in order to measure the
dielectric ac susceptibility $\chi \simeq \varepsilon =\varepsilon ^{\prime
}-i\varepsilon ^{\prime \prime }$ with an HP 4194 A impedance bridge with a
four wire probe and a signal level of 0.5 V/mm, between 200 Hz and 1 MHz,
and its mechanical equivalent, the dynamic compliance $s=s^{\prime
}-is^{\prime \prime }$, through the electrostatic excitation of the flexural
modes (1 and 13 kHz) in the linear regime, as described in Ref. %
\onlinecite{105}. All the measurements presented here were made on the same
sample.

The $\varepsilon $ curves are shown in Fig. 1, and agree with most of the
measurements of PLZT 9/65/35 in the literature,\cite{BKP99} with a frequency
dispersion typical of a freezing process at some temperature below the
susceptibility maxima, namely $T_{f}<340$~K.

The dynamic compliance is normalized to a reference value $s_{0}$. The
maxima of $s^{\prime }$ are shallower and shifted to lower temperature,
indicating that the dynamics of dipolar type (affecting $\varepsilon $ but
not $s$) and of quadrupolar type (affecting $s$) are different. In fact, $%
\varepsilon $ probes the fluctuations of electric dipoles, while $s$ of
local distortions; the latter, being symmetric strain tensors, are
quadrupoles.\cite{LB1} For the present purposes it is sufficient to note
that the quadrupolar dynamics exhibits an even lower freezing temperature.
The curves in Fig. 1 are the references against which aging and memory
curves will be compared.

Figure 2 shows the memory effect in both the susceptibilities, after aging 1
day at 298~K. Before the aging experiments, the sample was brought to a
reference state by heating to $\gtrsim 550$~K and cooling at 1.5~K/min to
the first aging temperature. The closed symbols are measured on cooling, and
the open symbols after subsequent heating. The reference curves measured
during continuous cooling have been subtracted from the data, and the
heating curves are multiplied by a factor $F$ in order to overlap with those
on cooling; $F^{-1}$ provides a measure of the memory effect, which turns
out to be almost the same in the anelastic (63\%) and dielectric (69\%)
cases. We also measured the isothermal decays of the susceptibilities during
aging at and above room temperature, but did not find any major difference
between the dielectric and elastic responses.

A run with multiple aging stages (24~h each) at 374, 349 and 324~K is
presented in Fig. 3; notice that the first two aging temperatures are
certainly above $T_{f}$. The dips on heating are better seen after dividing
by the reference curves; this is shown for the real part of the dielectric
permittivity in Fig. 4. The memory is almost complete for the last aging
stage at 324~K, while the other two stages are only partially retraced.
Similar effects are found also in the compliance curves (not shown here).

The degree of memory can be estimated by fitting the three dips with
gaussians, and defining the degree of memory as the ratio of the amplitude
of these gaussians to the lowest values reached during aging. In this
manner, one finds 90\%, 50\% and $\lesssim 80\%$ memory of the three aging
stages, starting from the lowest temperature. The gaussian form has been
adopted, since it fits very well the heating curve after a single aging
stage, as in Fig. 2; yet, the effects of multiple agings may be
non-additive, and, moreover, the gaussians are broader at higher
temperature. The frequency dependence of the dips at lower temperature is
due to differences in aging and not in memory, and the temperature
dependence of aging is in agreement with previous results.\cite{CCW01}


In traditional spin-glasses, aging and memory effects are found below the
glass transition temperature $T_{g}$, and their presence is even considered
a signature of the spin glass state (see e.g. Ref. \onlinecite{NMN00}). This
is true for both experiment and for various models of the spin glass state,
although it may be debated whether a true transition temperature exists and
is experimentally accessible, or rather it is hindered by the exceedingly
slow dynamics on approaching it. In the hierarchical scenario derived from
the Sherrington and Kirkpatrick model, the roughening of the potential
landscape, held responsible for these out-of equilibrium phenomena, starts
below $T_{g}$,\cite{MPP01,BM80} and the number of metastable minima
increases extremely fast with the reduced temperature $t=\left(
T_{g}-T\right) /T_{g}$.\cite{LOH91,BM80} In addition, the heights of the
barriers between the metastable states have been experimentally shown to
increase with cooling below $T_{g}$.\cite{MPP01} Similarly, in models
derived from the droplet picture\cite{FH86} and adapted to explain memory,%
\cite{BDH01} the height of the barriers between the metastable states
vanishes at $T_{g}$.

The remarkable feature of the results in Fig. 3 is that {\it the memory of
multiple aging stages in PLZT is retained also above the temperature of
freezing of the polar fluctuations}. This is at variance with other glassy
systems, including canonical Ising and Heisenberg spin glasses\cite{DVB01}
and dipolar glasses,\cite{ABC98} where non ergodic phenomena occur only
below a glass transition.\cite{ABC98}

The definition of a glass transition temperature in PLZT is not obvious,
but, by analogy with other model and real spin glass systems, it should be
very close to and in any case below the temperature at which the relevant
degrees of freedom appear to freeze. For relaxor ferroelectrics, the
polarization dynamics is the relevant one, and is probed by the dielectric
susceptibility, so that we can state that $T_{g}$ is below the temperature
of the maxima of the ac susceptibility. To be more quantitative, we may
adopt the usual argument that, at the maximum, the susceptibility is probing
fluctuations with characteristic relaxation time $\tau \sim \omega ^{-1}$,
and fit the temperature $T_{m}\left( \omega \right) $ of the susceptibility
maximum with the Vogel-Fulcher expression $\omega =\tau ^{-1}\sim \tau
_{0}^{-1}\exp \left[ -E/\left( T_{m}-T_{\text{VF}}\right) \right] $. Under
these naive assumptions, the maxima of $\varepsilon ^{\prime }\left( \omega
,T\right) $ in Fig. 1 indicate that the relaxation time diverges at $T_{%
\text{VF}}=320.5$~K with $\tau _{0}=2.1\times 10^{-11}$~s and $E=449~$K. On
the other hand, within a scaling approach\cite{DVB01,Myd93} with $\tau \sim
\left( \frac{T-T_{g}}{T_{g}}\right) ^{-z\nu }$, a fit of $T_{m}\left( \omega
\right) $ yields $T_{g}=337$~K and $z\nu =6.8$. The data of Fig. 1 do not
allow the two $\tau \left( T\right) $ laws to be distinguished, but
certainly indicate that the system cannot be considered frozen above $337$%
~K. One can compare the frequency dependence of the susceptibility maximum
with that of spin-glasses\cite{Myd93} and superparamagnets,\cite{FTD86}
finding that both the Vogel-Fulcher and the dynamical scaling parameters are
typical of the first class of materials: $E/k_{\text{B}}T_{\text{VF}}=1.4$
is between the values 0.85 and 2 found in the canonical spin glasses Cu$%
_{1-x}$Mn$_{x}$ and Eu$_{1-x}$Gd$_{x}$, while $E/k_{\text{B}}T_{\text{VF}}=11
$ for superparamagnets based on Fe particles; $z\nu =6.8$ is within the
range 4-8 found for spin glasses whereas superparamagnets may have $z\nu $
as high as 40.

Another signature of the onset of a non ergodic state is the splitting
between the field cooled (FC) and zero field cooled (ZFC) susceptibilities
or of the magnetization for magnetic spin glasses and polarization for
relaxor ferroelectrics. In PLZT 9/65/35 this occurs around 325~K, in
correspondence with the maximum of the static susceptibility,\cite{KFP99}
and in agreement with the above estimates.

On the other hand, dielectric measurements over an extremely broad range of
frequencies in the relaxor ferroelectrics PMN and PLZT,\cite{KFP99} allow
the distribution function of the relaxation times to be measured, and
suggest the definition of an even lower freezing temperature, namely the
temperature at which the longest relaxation time $\tau _{\max }$ diverges.%
\cite{KBP00} Below the Burns temperature $T_{B}\simeq 627$~K, there is a
strong broadening of the dielectric spectra with $\tau _{\max }$ following
the Vogel-Fulcher law with $T_{\text{VF}}\simeq 230$~K\cite{KBP00} or 250~K%
\cite{KFP99} in PLZT 9/65/35, depending on the type of analysis.

Summarizing, in PLZT\ one can identify a characteristic temperature for the
freezing of the polarization in different ways, but in any case, as
expected, below the temperature $T_{m}$ of the maxima of the dynamic
dielectric susceptibility, below 340~K in the present case. We therefore
expect a glass-like state with proliferation of metastable states below $%
T_{m}$, as usual in spin glasses, and it is striking to find that the memory
of multiple aging stages extends above that temperature. One could consider
other degrees of freedom, freezing at $T>T_{m}$, as responsible for the
formation of a glassy state and therefore of multiple memory phenomena. In a
relaxor ferroelectric like PLZT the other relevant degrees of freedom are
the elastic distortions (quadrupoles), probed by the elastic compliance $%
s\left( \omega ,T\right) $; however, the maxima of $s$ are at even lower
temperature than those of $\varepsilon $ (see Fig. 1), indicating that the
quadrupolar dynamics is faster, and is not a possible source of hidden
freezing before the dipolar freezing.

Another possible source of aging is the migration or reorientation of
defects and free charges, and is well known as the phenomenon of clamping of
the domain walls in ferroelectrics.\cite{RA93} In fact, aging has recently
been reported to occur also above the temperature of the dielectric maximum\
in PLZT,\cite{CCW01} and the phenomenology, more complex than in the other
relaxor ferroelectric PMN-PT, has been assigned to multiple aging
mechanisms, possibly including the slow motion of defects. It is in fact
conceivable that the slow rearrangement of defects may cause memory, if
during aging they leave a template defect configuration that remains frozen
during cooling, and induces the same highly pinned configuration on
subsequent heating. Such a mechanism, however, would work with any
ferroelectric, and could not produce the imprinting of several agings. In
fact, it requires that the defect configuration remains frozen at
temperatures below the first aging, and therefore cannot account for the
imprinting of the subsequent agings at lower temperatures. The present
results, instead, demonstrate the imprinting of {\it several} aging stages
above $T_{f}$.


Other perovskite systems presenting electric polarization freezing, aging
and memory are the so-called dipolar glasses, like KLT or KTN. The
non-equilibrium effects in these systems are at least an order of magnitude
smaller than those found here, but they have been extensively studied,\cite
{ABC00} and the aging part of susceptibility has been attributed to the
domain walls reconformation, when the domains slowly evolve toward the
equilibrium size.\cite{ABC00}

The same ideas have been used to explain the memory of multiple aging stages
in spin glasses by Bouchaud {\it et al}.,\cite{BDH01} who proposed a
qualitative picture that derives from the droplet model of spin glasses, but
satisfies also the hierarchical scenario. The main ideas are: {\it i)} the
time necessary for a reconformation of a domain or domain wall is $\tau \sim
\tau _{0}\exp \left( U/T\right) $; {\it ii)} the barrier for reconformations
over a length scale $l$ is $U\sim \Upsilon l^{\theta }$ with $\theta >0$; 
{\it iii) }the barrier vanishes at the glass transition temperature $T_{g}$, 
$\Upsilon =\Upsilon _{0}\left| 1-T/T_{g}\right| ^{-\nu }$. In this manner,
particularly in virtue of hypothesis ({\it iii), }there is a strong
separation of the time scales necessary for reconformations over different
lengths and temperatures. Therefore, at each temperature $T$ the system is
in equilibrium over a length scale $l<l\left( T\right) $, but almost frozen
over $l>l\left( T\right) $. Aging at $T_{1}$ produces reconformations over $%
l\le l\left( T_{1}\right) $, while aging at $T_{2}<T_{1}$ is effective only
at $l\le l\left( T_{2}\right) $; if the scale separation between $T_{1}$and $%
T_{2}$ is sufficient, then, after coming back at $T_{1}$, the system
restarts aging where it was left, except for a very fast initial
equilibration of the faster degrees of freedom that had changed during aging
at $T_{2}$. The reasoning can be extended to several aging temperatures $%
T_{n}$, if there is sufficient separation in the time and therefore length
scales, and hypothesis ({\it iii)} has been found essential to obtain such a
strong separation.\cite{BDH01} If the above scenario applies also to PLZT,
then the present data suggest that the hypothesis of vanishing barriers at
the freezing temperature may be lifted, and still multiple memory effects
are observable, although partially. The ground for supposing vanishing
barriers at $T_{g}$ in the domain wall scenario,\cite{BDH01} is the analogy
with the ferromagnetic case, where the wall stiffness $\Upsilon $ vanishes
at the ferromagnetic transition. In the case of PLZT and other relaxor
ferroelectrics, the onset temperature of the ferroelectric correlations is
the Burns temperature $T_{B}\gg T_{g}$ and the barriers for domain wall
motion are expected to vanish at that temperature.

The degree of (single aging) memory has been found to depend on temperature,%
\cite{CCW01} and it may be considered as a measure of the separation between
the length scales involved in aging, or equivalently of the degree of
hierarchization of the system. The heating curves in Fig. 3 indicate that
memory does not disappear above $T_{g}$, but gradually fades out above the
susceptibility maximum. It remains to be clarified whether the aging and
memory effects above $T_{f}$ are still evidence of a hierarchical
organization of metastable states, and what would its origin be.

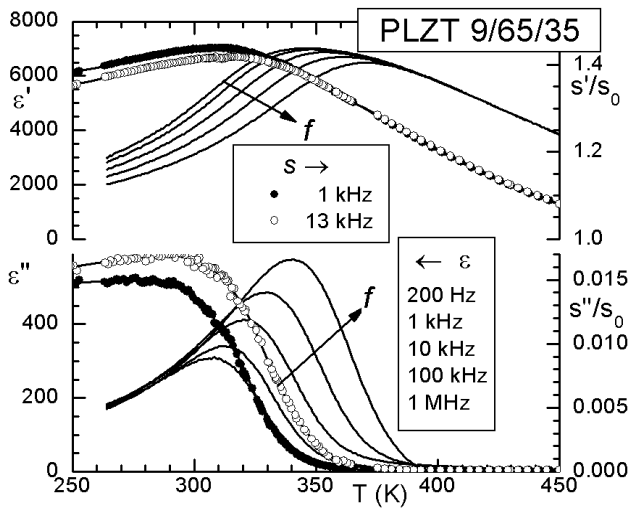
\begin{figure}[tbp]
\caption{Real and imaginary parts of the dielectric susceptibility (lines,
left hand ordinates) and elastic compliance (circles, right hand ordinates)
of PLZT 9/65/35.}
\end{figure}

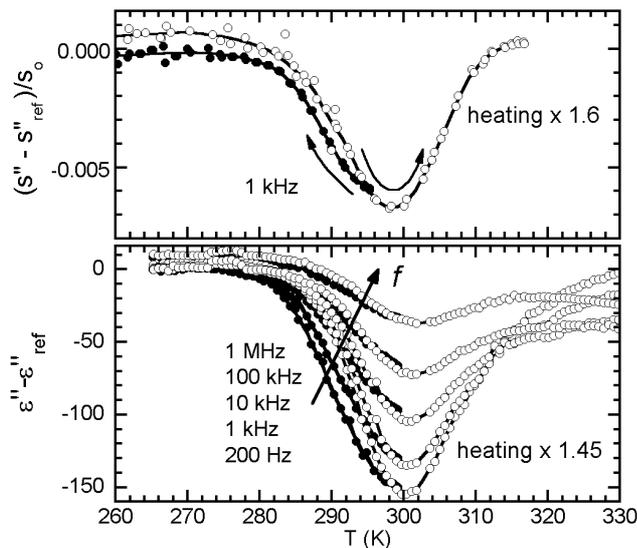
\begin{figure}[tbp]
\caption{Rejuvenation on cooling (closed symbols) after aging 1 day at
298~K and memory on heating (open symbols) in the elastic compliance
(top panel) and dielectric permittivity (bottom panel). The reference curves
of Fig. 1 have been subtracted.}
\end{figure}

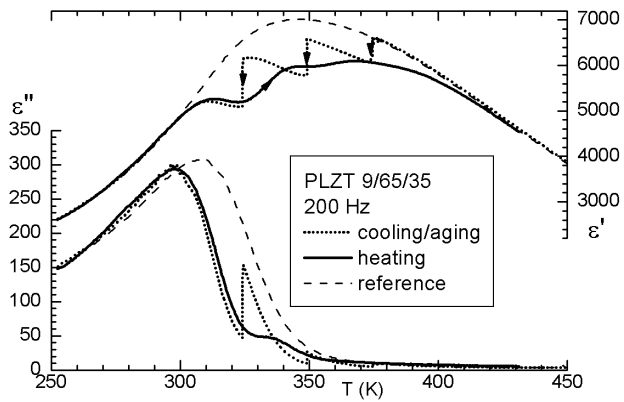
\begin{figure}[tbp]
\caption{Dielectric permittivity at 200~Hz during multiple aging stages
(24~h each) and subsequent heating curves with memory, compared with the
reference curve measured on continuous cooling.}
\end{figure}

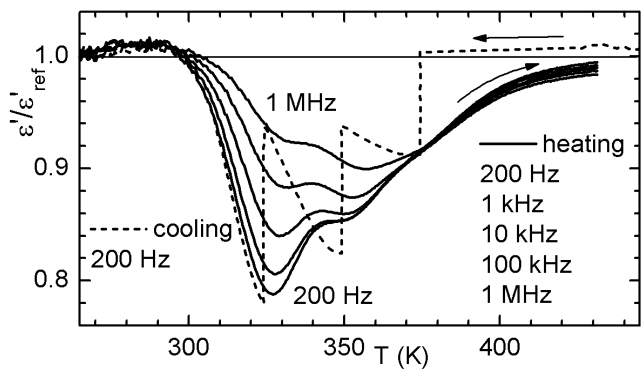
\begin{figure}[tbp]
\caption{Relative change of the real part of the permittivity measured with
the temperature protocol of Fig. 3 with respect to the reference.}
\end{figure}


\begin{references}
\bibitem{NMN00}  D.N.H. Nam {\it et al.}, Phys. Rev. B {\bf 62}, 8989 (2000).

\bibitem{DVB01}  V. Dupuis {\it et al.}, Phys. Rev. B {\bf 64}, 174204
(2001).

\bibitem{MPP01}  R. Mulet {\it et al.}, Phys. Rev. B {\bf 63}, 184438 (2001).

\bibitem{BDH01}  J.-P. Bouchaud {\it et al.}, Phys. Rev. B {\bf 65}, 24439
(2001).

\bibitem{VHO96}  E. Vincent {\it et al.}, in {\it Complex Behavior of Glassy
Systems}, ed. by M. Rubi, (Springer-Verlag, Berlin, 1997).

\bibitem{CCW01}  E.V. Colla {\it et al.}, Phys. Rev. B {\bf 63}, 134107
(2001).

\bibitem{KB02b}  O. Kircher and R. B\"{o}hmer, Eur. Phys. J. B {\bf 26}, 329
(2002).

\bibitem{KBP00}  S. Kamba {\it et al.}, J. Phys.: Condens. Matter {\bf 12},
497 (2000).

\bibitem{PB99}  R. Pirc and R. Blinc, Phys. Rev. B {\bf 60}, 13470 (1999).

\bibitem{SK75}  D. Sherrington and S. Kirkpatrick, Phys. Rev. Lett. {\bf 35}%
, 1792 (1975).

\bibitem{LOH91}  M. Lederman {\it et al.}, Phys. Rev. B {\bf 44}, 7403
(1991).

\bibitem{VXP93}  D. Viehland {\it et al.}, J. Appl. Phys. {\bf 74}, 7454
(1993).

\bibitem{105}  F. Cordero {\it et al.}, Ferroelectrics {\bf 290}, 141 (2003).

\bibitem{BKP99}  V. Bobnar {\it et al.}, Europhys. Lett. {\bf 48}, 326
(1999).

\bibitem{LB1}  G. Leibfried and N. Breuer, {\it Point Defects in Metals I}.
ed. by M. Levy, (Springer, Berlin, 1978).

\bibitem{BM80}  A.J. Bray and M.A. Moore, J. Phys. C: Solid State Phys. {\bf %
13}, L469 (1980).

\bibitem{FH86}  D.S. Fisher and D.A. Huse, Phys. Rev. Lett. {\bf 56}, 1601
(1986).

\bibitem{ABC98}  F. Alberici-Kious {\it et al.}, Phys. Rev. Lett. {\bf 81},
4987 (1998).

\bibitem{Myd93}  J.A. Mydosh, {\it Spin glasses. An experimental introduction%
}. (Taylor \& Francis, London, 1993).

\bibitem{FTD86}  D. Fiorani {\it et al.}, J. Phys. C {\bf 19}, 5495 (1986).

\bibitem{KFP99}  Z. Kutnjak {\it et al.}, Phys. Rev. B {\bf 59}, 294 (1999).

\bibitem{RA93}  U. Robels and G. Arlt, J. Appl. Phys. {\bf 73}, 3454 (1993).

\bibitem{ABC00}  F. Alberici-Kious {\it et al.}, Phys. Rev. B {\bf 62},
14766 (2000).
\end{references}
\end{document}